\documentclass[twocolumn,aps,epsfig]{revtex4}
\usepackage{graphicx}

\usepackage{latexsym} 
\usepackage{mathptm}
\usepackage{epsfig}
\usepackage{times}
\usepackage{slashbox}

\def\beq{\begin{equation}}
\def\eeq{\end{equation}}
\def\beqn{\begin{eqnarray}}
\def\eeqn{\end{eqnarray}}
 
\begin{document}

\title{A Bi-Metric Theory with Exchange Symmetry}
\author{S.~Hossenfelder}
\email{sabine@perimeterinstitute.ca}

\affiliation{Perimeter Institute for Theoretical Physics\\ 
31 Caroline St. N, Waterloo Ontario, N2L 2Y5, Canada}

\date{\today}

\begin{abstract} 
We propose an extension of General Relativity with two different metrics. To each metric we define a
Levi-Cevita connection and a curvature tensor. We then consider two types of fields, each of which moves 
according to one of the metrics and its connection. To obtain the field equations for the second
metric we impose an exchange symmetry on the action. As a consequence of this ansatz,
additional source terms for Einstein's field equations are generated. We discuss the properties of these additional
fields, and consider the examples of the Schwarzschild solution, and the Friedmann-Robertson-Walker 
metric. 
\end{abstract}

\maketitle
 
During the last decades, experimental achievements in astrophysics provided us with new insights about
our universe. The more precise our observations have become, the more obvious also the insufficiency of
our understanding have become. Today's research in cosmology is accompanied by the group of
cosmological problems, which strongly indicate that our knowledge about the universe is incomplete. 
Most importantly, it is microscopic explanations for dark matter and dark energy that we are lacking. A lot
of effort has been invested into studies of fields with unusual equations of states which can account
for one or the other constituent.

In this paper we propose an extension of General Relativity with fields that experience space-time to
have a metric ${\bf \underline h}$ different from our usual one ${\bf g}$. This causes additional source terms to 
Einstein's field equations that have properties unlike those of our standard matter. The new
fields do not move according to the Levi-Cevita connection, but according to a non-metric, torsion-free connection derived from the second metric ${\bf \underline h}$. To obtain the equations of motion for the second metric, we
propose a symmetry between both types of matter and the according metrics.  
  
Different versions of bi-metric theories and their potential importance to explaining observational
evidence have previously been investigated in \cite{linde1,linde2,Drummond:2001rj,Moffat:2002kj}, and the
approaches in references \cite{Bondi,Quiros:2004ge,Borde:2001fk,Davies:2002bg,Ray:2002ts,Rosenberg:2000cv,Torres:1998cu,Zhuravlev:2004vd,Faraoni:2004is,Henry-Couannier:2004vt,Moffat:2005ip,Kaplan:2005rr,Hossenfelder:2005gu,Nickner:2006xh} study similar symmetry considerations. 

This paper is organized as follows:  The general setup with the two metrics and two types of fields is introduced 
in section \ref{setup}. In the section \ref{con}, we define the connections, and in section \ref{act} we construct the action for the new sort of fields and couple it to General Relativity (GR). In section \ref{ex} we use the exchange symmetry to obtain the complete set of equations including those for the second metric, and then investigate the example of the
Schwarzschild solution, and the Friedmann-Robertson-Walker metric in \ref{example1} and \ref{example2}. After a discussion of
the scenario and its possible observable consequences in \ref{disc}, we conclude in section \ref{conc}.

Throughout this paper we use the convention $c=\hbar=1$. The signature of the metric is $(-1,1,1,1)$. Small
Greek indices are space-time indices and run from 0 to 3.

\section{A Bi-Metric Theory}
\label{setup}

We consider a bi-metric theory with metrics ${\bf g}$ and ${\bf \underline h}$ of Lorentzian signature that define two different ways of measuring angles, distances and volumes on a manifold $M$. Changing from ${\bf g}$ to ${\bf \underline h}$ changes the map from the tangential space, $TM$, to the co-tangential space, $TM^*$. Since we then have two ways to raise and lower indices, 
we will use a notation with two types of coordinate-indices so we do not spoil the Ricci-calculus. For this, we will denote coordinate indices that are raised and lowered by ${\bf \underline h}$ with underlines. In case objects do not carry indices we underline them in total. 

We will further introduce two sorts of matter on $M$: one that moves according to the usual\footnote{Since we will later employ a symmetry between both metrics, calling one the `usual' is merely a convention.} metric ${\bf g}$ and the measure it implies, the other one that moves according to the other metric ${\bf \underline h}$. We will refer to these fields as $g$-fields and $h$-fields, respectively; the equations of motions will be specified in section \ref{act}.  Related, we consider two types of observers on our spacetime, the one made up of $h$-fields who measures with the metric ${\bf \underline h}$, the other one made of $g$-fields who measures with metric ${\bf g}$. They will have to set their observations in relation to each other in a consistent way, much like different observers in Special Relativity. 

Such, we have on the one hand the $h$-observer who sees a $g$-field with unusual behavior, and on the other hand the $g$-observer who thinks this field to be perfectly normal matter. The same situation applies for $g$ and $h$ exchanged. To take this into account we will consider a map $P_{\underline h}$, which is an automorphism on the tensor-bundle, and which maps $h$-fields as the $h$-observer sees them to $h$-fields as the $g$-observer sees them. Similarly, we have a map $P_g$, which maps $g$-fields as the $g$-observer sees them to $g$ fields as the $h$-observer sees them. These maps conserve the tensor structure of objects, i.e. a tensor of rank $(r,s)$ is mapped to a tensor of rank $(r,s)$, and they are linear in the field's components. Most importantly, they assign a two-tensor $h_{\kappa \nu}$ to the metric ${\bf \underline h}$, which we will denote by ${\bf h}$, and a two-tensor $g_{\underline {\kappa \nu}}$ to the metric ${\bf g}$, which will be denoted by ${\underline{\bf g}}$: 
\beqn
{\underline{\bf g}} = P_g({\bf g}) \quad, \quad {\bf h} = P_{\underline h}({\underline{\bf h}}) \quad. \label{ps}
\eeqn
Since these maps are linear, we can write in components
\beqn
 \left[P_{\underline h}\right]^{\underline \nu}_{\;\;\nu} 
\left[P_{\underline h}\right]^{\underline \kappa}_{\;\;\kappa} h_{\underline \nu \underline \kappa} &=& h_{\nu \kappa} \\ 
\left[P_g\right]^{\nu}_{\;\;\underline \nu} \left[P_g\right]^{\kappa}_{\;\;\underline \kappa} g_{\nu \kappa} &=& g_{\underline \nu \underline \kappa}\quad. 
\eeqn 
We will in the following refer to these maps as `pull-overs'. One has to be careful however when pulling elements 
over from the one observer to the other. Since 
the components of pulled over tensors are mapped from the tangential into the cotangential space with a different metric, 
the notation of the inverse does no longer match with the use of indices. I.e. the pull-over $P^{-1}_{\underline h}$ of $h^{\underline \nu 
\underline \kappa}$ is not $h_{\kappa \nu} g^{\kappa \epsilon} g^{\nu \lambda}$ (where $h_{\kappa \nu}$ is the pull-over of $h_{\underline{\kappa \nu}}$) but rather a tensor ${\bf h}^{-1}$ with the property that $[h^{-1}]^{\epsilon \lambda} h_{\epsilon \kappa} = \delta^{\lambda}_{\;\; \kappa}$. We will try to avoid this by not using the pull-overs 
as long as not absolutely necessary (and we will see that we can rather successfully do so).

The introduction of these pull-overs is an assumption. Its usefulness lies in enabling us to now chose different coordinate
systems for the $g$ and $h$-observer. We could for example apply a change of coordinates only on the ${\underline{\bf h}}$ metric. This would imply an according change in the pull-over (a multiplication with the inverse of the coordinate transformation), but it would not make it necessary to also change the coordinate system the $g$-observer has chosen for 
his description. The most obvious choice for the pull-over would just be the identity. We will however see later that 
we can not fix these pull-overs, but that they have to be determined from the field equations, and can not in general be
chosen to be the identity.

We further define a map ${\bf a}$ that transforms the one metric into the pull-over of the other  
\beqn
g_{\epsilon \lambda}  = a_{\epsilon}^{\;\;\nu} a_{\lambda}^{\;\;\kappa} ~ h_{\nu \kappa}   \label{haga} \quad.
\eeqn
Since both ${\bf g}$ and ${\bf h}$ are symmetric, ${\bf a}$ is not completely determined by (\ref{haga}). We
fix the remaining six degrees of freedom (dof) by requiring it to be symmetric, i.e. $g^{\kappa \nu} a^{\epsilon}_{\;\;\nu} = a^{\epsilon \kappa} = a^{\kappa \epsilon}$. We can pull over ${\bf a}$ by
\beqn
a_{\underline \epsilon}^{\;\;\underline \nu} = 
\left[P_{g}\right]^{\epsilon}_{\;\;\underline \epsilon} a_{\epsilon}^{\;\; \nu}\left[ P_{\underline h} \right]^{\underline \nu}_{\;\;\nu}  \quad,
\eeqn
which then gives the relation
\beqn
g_{\underline{\epsilon \lambda}} = a_{\underline \epsilon}^{\;\; \underline \nu} a_{\underline \lambda}^{\;\; \underline \kappa} ~ h_{\underline{\nu \kappa}} \label{hagau}  \quad.
\eeqn
This pulled over quantity is also required to be symmetric. It is further useful to define a combination of ${\bf a}$ and the pull-overs that maps ${\bf g}$ to ${\bf \underline h}$ via
\beqn
a_{\epsilon}^{\;\;\underline \nu}  &=& a_{\epsilon}^{\;\;\nu} [P_{\underline h}]^{\underline \nu}_{\;\;\nu} \quad, \label{ua} \\
  g_{\epsilon \lambda} &=& a_{\epsilon}^{\;\;\underline \nu} a_{\lambda}^{\;\;\underline \kappa} ~ h_{\underline{\nu\kappa}}  \quad. \label{gauha}
\eeqn
And by raising and lowering some indices we also have
\beqn
g_{\epsilon \lambda} = a_{\epsilon}^{\;\;\underline \nu} a_{\lambda \underline \nu}~,~ 
 h_{{\underline{\nu \kappa}}} = a^{\epsilon}_{\;\;\underline{\nu}} a_{\epsilon \underline{\kappa}}  
  \quad.
\eeqn
In this formulation, the introduced map ${\bf a}$ is a convenience and not a dynamical field, since it is defined 
by relating ${\bf g}$ to $P_{\underline h}({\bf \underline h})$. The dynamical 
quantities are ${\bf g}$ and ${\bf{\underline h}}$, as well as the both pull-overs $P_{\underline h}$ and $P_g$, and possible additional matter and gauge fields. That means for the variation we have to keep ${\bf a}$ fixed. We specify the properties of the field under variation by requiring
\beqn
\delta a^{\nu\kappa} = 0 \label{vara} \quad.
\eeqn
We will see later that this requirement makes for an interesting scenario as it results in quite 
unusual properties of the $h$-fields. One can read this off already from Eq. (\ref{haga}). Demanding (\ref{vara}) to hold will imply that it is the inverse of the second metric that behaves under variation like the usual metric. Unlike the
problems that one would run into by just using the inverse metric as a second metric (because one had to treat a covariant field as a contravariant one) the approach proposed here by introducing a field that only behaves under variation as the inverse metric, is manifestly covariant.   

\section{Connections}
\label{con}

To ${\bf g}$ one can define a Levi-Cevita connection in the usual way that we will denote as ${}^{(g)}\nabla$. Similarly, one can define a Levi-Cevita connection to ${\bf \underline h}$, and we will denote this connection as ${}^{({\underline h})}\underline \nabla$. To both metrics with their connections, one can construct the curvature tensor, the Ricci tensor, and the curvature scalar, that we will denote as ${}^{(g)} R$ and ${}^{({\underline h})} R$, respectively. In detail one has
\beqn
{}^{(g)} \Gamma^\epsilon_{\;\nu\kappa} &=& g^{\epsilon \alpha } {}^{(g)}\Gamma_{\alpha \nu\kappa} \quad,\\
 {}^{(g)}\Gamma_{\alpha \nu\kappa} &=& \frac{1}{2}
\left( \partial_{\nu} g_{\alpha \kappa} + \partial_{\kappa} g_{\alpha \nu} - \partial_\alpha g_{\nu \kappa} \right) \quad,\\
{}^{(h)} \underline{\Gamma}^{\underline \epsilon}_{\;\underline \nu \underline \kappa} &=& h^{\underline \epsilon \underline \alpha} 
{}^{({\underline h})} \underline \Gamma_{\underline \alpha \underline \nu \underline \kappa}\quad,\\
{}^{({\underline h})} \underline \Gamma_{\underline{\alpha\nu\kappa}} &=& \frac{1}{2} (\partial_{\underline \nu} 
h_{\underline{\alpha\kappa}} + \partial_{\underline \kappa} h_{\underline{\alpha\nu}}-\partial_{\underline \alpha} h_{\underline{\nu\kappa}}) \quad.
\eeqn
We can further define a pulled over derivative for the $h$-field, which we denote as ${}^{(h)}\nabla$, by requiring it to be a torsion-free but non-metric connection that preserves ${\bf h}$. From ${}^{(h)}\nabla {\bf h}=0$ we then find the connection coefficients to ${}^{(h)}\nabla$ to be
\beqn
{}^{({h})} \Gamma^{\epsilon}_{\;\; \nu \kappa} =  \frac{1}{2} [P_{\underline h}({\bf {\underline h}})]^{\epsilon \alpha}\left( \partial_{\nu} h_{\alpha \kappa} + \partial_{\kappa} h_{\alpha \nu} - \partial_\alpha h_{\nu \kappa} \right)  \label{connh}\quad.
\eeqn
As mentioned previously, one should keep in mind that $[P_{\underline h}({\bf {\underline h}})]^{\epsilon\alpha}$ is not $ h^{\epsilon\alpha}$ but the inverse of $h_{\epsilon\kappa}$, and further that the indices on ${}^{({h})} \Gamma$ are lowered and raised with ${\bf g}$. From the above one also finds for $h$, the determinant of $h_{\nu\kappa}$,
\beqn
\partial_\kappa h = 2 h {}^{(h)} \Gamma^{\nu}_{\;\;\nu\kappa} \quad. \label{detder}
\eeqn
One can do a similar construction to pull over ${}^{(g)}\nabla$ to obtain a derivative ${}^{({\underline g})}\underline \nabla$, which is not metric with respect to ${\bf{\underline h}}$ but preserves ${\bf \underline g}$, i.e. ${}^{(\underline g)}\underline \nabla {\bf \underline g} =0$. This construction of derivatives now 
puts a further requirement on the pull-overs because they have to be compatible with  the tensor structure. So, for an arbitrary tensor ${\bf A}$ we then have
\beqn
P_{\underline h}({}^{(\underline h)}\underline \nabla {\bf A}) &=& {}^{(h)}\nabla P_{\underline h}({\bf A}) \quad, \\
P_{g}({}^{(g)} \nabla {\bf A}) &=& {}^{(\underline g)}\underline \nabla P_{g}({\bf A}) \quad .
\eeqn
In components we had e.g. for $A^{\underline \lambda}_{\;\;\underline{\kappa \epsilon}}$  
\beqn
[P_{\underline h}({}^{(\underline h)}\underline \nabla_{\underline \nu} A^{\underline \lambda}_{\;\;\underline{\kappa \epsilon}})]^{\lambda}_{\;\; \nu \kappa \epsilon} =  {}^{(h)} \nabla_{\nu} [P_{\underline h}(A^{\underline \lambda}_{\;\; \underline{\kappa \epsilon}})]^{\lambda}_{\;\; {\kappa \epsilon}} \quad.
\eeqn
Since both connections as well as their pull-overs have to be torsion free, this implies the pull-overs have to be integrable and are generated by two vector fields $v$ and $w$ such that $[P_g]^{\nu}_{\;\;\underline \kappa} \equiv W^{\nu}_{\;\;\underline \kappa} = \partial_{\underline \kappa} w^{\nu}$, and $[P_{\underline h}]^{\underline \nu}_{\;\;\kappa} \equiv \underline V^{\underline \nu}_{\;\;\kappa} = \partial_\kappa v^{\underline \nu}$. The pull-overs thus carry four dof each, after requiring them to be torsion free and metric-compatible in the above described fashion.

However, we will for the variation not assume these requirements are already fulfilled, as this is in conflict with the
dof we need. As explained previously, the independent variables are ${\bf g}$ and $\underline{\bf h}$,
each of which has ten dof. Since the ${\bf a}$'s and ${\bf \underline a}$'s do not carry degrees of freedom, the pull-overs need to carry these ten dof since a conjunction of both, the ${\bf a}$'s and the pull-overs, relates ${\bf g}$ to ${\bf \underline h}$, as can be read off from Eq. (\ref{gauha}). Therefore, prior to the variation we can 
only assume the connections are torsion free, and take into account metric compatibility after the variation. For the usual connections, the variation over the connection together with the torsion-free-ness implies metric compatibility as usual. For the additional connections, we will here not explicitly add a term to the action to generate it but subsequently assume metric-compatibility since it seems to be a desirable feature (though the scenario could be considered in more generality).  
 
To summarize this section: Each metric defines its own Levi-Cevita connection, and after pulling them over these induce two non-metric connections, which will describe the motion the $g$-observer assigns to the $h$-fields and vice versa. 

\section{And Action}
\label{act}

Now let us add some physics. Consider we have an $h$-field that behaves not according to the usual Levi-Cevita connection, but according to the connection metric with respect to ${\underline{\bf h}}$, a field that feels angles and distances as defined by ${\underline{\bf h}}$ not ${\bf g}$. For a massless scalar $h$-field $\underline \phi$ the action that gives
the equations of motions as the $g$-observer sees them could look like 
\beqn
S = \int d^4  x \sqrt{- h}~ P_{\underline h}\left( h^{{\underline{\nu \kappa}}} ~ {}^{(\underline h)}\underline \nabla_{\underline{\kappa}} \underline \phi {}^{(\underline h)}\underline \nabla_{\underline{\nu}} \underline \phi \right) \quad,
\eeqn
where $h = {\rm det}(P_{\underline h}(\underline{{\bf h}}))$ so we have pulled over the determinant and the measure is appropriately invariant.  
   
For a scalar field the covariant derivative is of course just the partial one so it does not matter according to which metric the connection is metric, but in general this will not be the case. One can construct Lagrangians $\underline {\cal L}$ for other types of $h$-fields than scalars in a similar way by replacing the usual metric with the other one, and the usual Levi-Cevita connection with the one belonging to the other metric, and then pulling over. If it was just to make a scalar, one could consider the density weight to be $\sqrt{-g}$. However, the relevance of putting $\sqrt{-{h}}$ instead becomes apparent when one takes the variation over the field and its connection to obtain the equations of motion (eom). In order to convert the pull-over of a term ${}^{({\underline h})}\nabla^{\underline \nu} A_{\underline \nu}$ into a total derivative the prefactor needs to be $\sqrt{-h}$ not $\sqrt{-g}$, so it is compatible with the derivation used in the Lagrangian. For such a term then to vanish one uses equation (\ref{detder}), which guarantees the validity of Gauss's law. The resulting eom are then just 
\beqn
P_{\underline h} \left( {}^{({\underline h})}\underline \nabla^{\underline \alpha} {}^{({\underline h})}\underline \nabla_{\underline \alpha} \underline \phi \right) = 0 \quad, \label{eomupsi}
\eeqn
which is by definition of the pull-over identical to
\beqn
{}^{(h)} \nabla^{\underline \alpha} {}^{(h)}\nabla_{\underline \alpha} P_{\underline h}( \underline \phi) = 0\quad.
\eeqn
Since the pull-over is invertible, the eom (\ref{eomupsi}) are also equivalent to
\beqn
{}^{({\underline h})}\underline \nabla^{\underline \alpha} {}^{({\underline h})}\underline \nabla_{\underline \alpha} \underline \phi = 0
\quad,
\eeqn
which are the eom the $h$-observer would expect. Now we add such a field to GR:
\beqn
S = \int d^4 x \sqrt{-g} \left( {}^{(g)}R/8 \pi G + {\cal L} \right) +  \sqrt{- h} ~  P_{\underline h}( \underline {\cal L})  \quad. \label{action} 
\eeqn
Upon variation, the first two terms give just the standard, and the last term yields
\beqn
\int d^4 x \frac{\delta}{\delta h_{\nu \kappa}} \left( \sqrt{-h}~   P_{\underline h}(\underline {\cal L}) \right)  \delta h_{ \nu\kappa} \quad,
\eeqn
which we have to rewrite into a variation over $g_{\kappa \nu}$ so we can add the terms. We vary (see also Appendix A)
\beqn
  g_{\epsilon \lambda} h_{\kappa\nu} a^{\epsilon \kappa} a^{\mu\nu} = \delta^{\mu}_{\;\; \lambda}
\eeqn
and rewrite into
\beqn
\delta h_{\kappa \lambda} = - [a^{-1}]^\mu_{\;\;\kappa} [a^{-1}]^{\nu}_{\;\;\lambda} \delta g_{\mu \nu} \quad, \label{sign}
\eeqn
where ${\bf a}^{-1}$ is the inverse of ${\bf a}$
\beqn
[a^{-1}]_{\beta}^{\;\; \kappa} a_{\nu}^{\;\; \beta} = \delta_\nu^{\;\;\kappa} \quad.
\eeqn
We  put in some pull-overs and their inverse, and rewrite into $a_\kappa^{\;\; \underline \kappa}$ to make the symmetry more apparent. Then, we can add all terms together and obtain from the variation (see Appendix A) of the action the equations 
\beqn
{}^{(g)}R_{\kappa \nu} - \frac{1}{2} g_{\kappa \nu} {}^{(g)}R = 8 \pi G \left( T_{\kappa \nu} - \sqrt{\frac{h}{g}} a_\nu^{\;\;\underline \nu} a_\kappa^{\;\;\underline \kappa} \underline T_{\underline{\nu \kappa}} \right) \label{fe1}
\eeqn
with the sources
\beqn
T_{\mu \nu} &=& - \frac{1}{\sqrt{-g}} \frac{\delta {\cal L}}{\delta g^{\mu\nu}} + \frac{1}{2} g_{\mu \nu} {\cal L} \label{set1}
\\
\underline T_{\underline{\nu \kappa}} &=& - \frac{1}{\sqrt{-\underline{h}}} 
\frac{\delta \underline {\cal L}}{\delta 
h^{\underline{\nu \kappa}}} + \frac{1}{2} h_{\underline{\nu \kappa}} \underline {\cal L} \nonumber \\
&=& \left[ P_{\underline h}^{-1} \left( - \frac{1}{\sqrt{-{h}}} \frac{\delta \underline {\cal L}}{\delta 
h^{{\nu \kappa}}} + \frac{1}{2} h_{{\nu \kappa}} \underline {\cal L} \right) \right]_{\underline{\nu\kappa}} \quad, \label{set2}
\eeqn
where the last line is only to clarify how the rewriting of the variation from ${\bf h}$ to ${\underline {\bf h}}$  comes into play.

This change of sign we see appearing here in equation (\ref{sign}) does occur only for the gravitational stress-energy-tensor, i.e. the source term to Einstein's field equations since it is a consequence of taking the variation with
respect to the metric. It does not occur if one derives the kinetic energy momentum tensor (via Noether's theorem), neither does the Lagrangian have a negative kinetic energy  term. Since then the sum as well as the difference of both stress-energy-tensors is conserved, this means they are separately conserved; both fields interact only gravitationally. Thus, despite the presence of negative gravitational masses, there is no vacuum instability because the kinetic energy of both sorts of fields remains positive, is conserved as usual, and a pair production of negative and positive gravitational masses out of vacuum is not possible. 

\section{With Exchange Symmetry}
\label{ex}

In the previous section we have only considered the perspective of the $g$-observer and the field equations for
the ${\bf g}$ metric. We have however used that the fields ${\bf g}$ and ${\bf h}$ are not independent. For symmetry reasons, the independent variables should be 
${\bf g}$ and ${\underline{\bf h}}$, as well as the two pull-overs, with which we further obtain ${\underline{\bf g}}$ and ${\bf h}$ via (\ref{ps}). ${\bf h}$ is then however related to ${\bf g}$ via eq. (\ref{haga}), and ${\bf{\underline g}}$ to ${\underline{\bf h}}$ via eq. (\ref{hagau}). 

Based on this consideration, we add matter fields to GR: a $g$-field $\psi$, and an $h$-field $\underline \phi$, and request the action be symmetric under exchange of ${\bf g}$ with ${\bf \underline h}$, and exchange of $g$-fields with $h$-fields. This way we obtain 
\beqn
S &=& \int d^4x \sqrt{-g} \left( {}^{(g)} R/8 \pi G + {\cal L}(\psi) \right) + \sqrt{-h} P_{\underline h}(\underline {\cal L}(\underline \phi)) \nonumber \\
&+& \int d^4  x \sqrt{- \underline h} \left({}^{(\underline h)}R/8 \pi G + \underline {\cal L}(\underline \phi) \right) + \sqrt{-\underline g} P_g({\cal L}(\psi)) \quad, 
\eeqn
where the first two terms are varied with respect to ${\bf g}$ using eq. (\ref{haga}) as done in the previous section, the last two terms with respect to $\bf{\underline h}$ using eq. (\ref{hagau}), and one should keep in mind that $h = {\rm{det}}(P_{\underline h}({\bf \underline h})) \neq \underline h$ and ${\underline g} = {\rm{det}}(P_{g}({\bf g})) \neq g$. The eom for the matter fields are the usual ones and their pull-overs, and the missing field equations for the second metric 
take the form
\beqn
{}^{(h)}R_{\underline{\nu \kappa}} - \frac{1}{2} h_{\underline {\nu \kappa}} {}^{({\underline h})}R = 8 \pi G \left( \underline T_{\underline{\nu \kappa}} - \sqrt{\frac{\underline g}{{\underline{h}}}} a^{\kappa}_{\;\underline{\kappa}} a^{\nu}_{\;\underline{\nu}} T_{\kappa \nu} \right) \quad, \label{fe2}
\eeqn
with the previously defined stress-energy tensors from eqs. (\ref{set1}) and (\ref{set2}).  Since equation (\ref{fe1}) contains $h$ rather than $\underline h$ and equation (\ref{fe2}) contains $\underline g$ rather than $g$, the pull-overs are necessary ingredients. We can make this more apparent by explicitly putting them into the equations:
\beqn
{}^{(g)}R_{\kappa \nu} - \frac{1}{2} g_{\kappa \nu} {}^{(g)}R &=& T_{\kappa \nu} - {\underline V} \sqrt{\frac{ \underline h }{g}} a_\nu^{\;\;\underline{\nu}} a_\kappa^{\;\;\underline { \kappa}} \underline T_{\underline{\nu \kappa}} \label{fe1b} \\
{}^{({\underline h})}R_{\underline{\nu \kappa}} - \frac{1}{2} h_{\underline{\nu \kappa}} {}^{(\underline h)}R &=& \underline T_{\underline{\nu \kappa}} - W \sqrt{\frac{ g}{{\underline{h}}} }a^\kappa_{\;\underline{\kappa}} a^\nu_{\;\underline \nu} T_{\kappa \nu} \quad, \label{fe2b}
\eeqn
where ${\underline V}$ is the determinant of $P_{\underline h}$, and $W$ is the determinant of $P_g$. (Or, to be more precise their absolute values since the volume element is positive.) Note that these equations would not be invariant under coordinate transformations for each of the observers separately without the pull-overs, since the factor $\underline h/g$ was not an invariant in this case.
 
The virtue of doing this is that we can now chose the coordinate systems for ${\bf g}$ and $\bf \underline h$ separately which seems to be natural. For example, there is no reason to expect that the coordinate system that one observer would consider `free-falling' will agree with the other one's, since they both move according to different connections. Thus it will not in general be clear what a gauge condition on the one metric does to the other one, and it seems more useful to export this lack of knowledge into the pull-overs.  

The field equations (\ref{fe1b}) and (\ref{fe2b}) need to fulfill the contracted Bianchi identities with respect to the 
matching Levi-Cevita connection. Since the stress-energy tensor of the $g$-field is already covariantly conserved with respect to ${}^{(g)}\nabla$, and the stress-energy tensor of the $g$-field is similarly covariantly conserved with respect to ${}^{(\underline h)} \underline \nabla$, this leaves us with each four equations for the additional source terms constituted of the fields behaving with respect to the non-metric connection. After assuming metric-compatibility of the pulled-over connections as explained in section \ref{con}, these each four equations constrain the remaining four dof in the pull-overs that appear in these terms. 

One should note however that the pull-overs will need additional initial conditions, and unless one imposes further symmetry requirements esp. their determinants can be multiplied by an arbitrary constant. One can also interpret this as there a priori being no way of telling whether the coupling
between the two types of fields and gravity is equally strong or has an additional pre-factor. However, instead of introducing additional coupling constants we will leave the constants in the pull-overs and treat them as parameters of the model that, ideally, have to be determined by observational constraints. Eqs. (\ref{fe1b}) reduce to the standard field equations in the limit where the energy density of the $h$-fields is very small and/or the determinant of the pull-over is small such that the coupling is very weak. 

To summarize this section: We have 10 components for each ${\bf g}$ and ${\underline{\bf h}}$. Equations (\ref{fe1b}) and (\ref{fe2b}) provide each 10 equations that are related by the (contracted) Bianchi identities. These two times 4 equations fix the two times 4 degrees of freedom left in $P_g$ and $P_{\underline h}$ after requiring metric-compatibility of the pulled-over connections, which leaves us as usual with 4 degrees of freedom to chose the coordinate systems for each metric.

\section{Example I: The Schwarzschild Metric}
\label{example1}

To obtain a better understanding of the workings, we consider the case with $\underline {\bf T} \equiv 0$, and with only a spherically symmetric source of usual $g$-fields  outside of which there is vacuum and we thus have the Schwarzschild solution for ${\bf g}$. This solution has one free constant, $M$, that is the integral over the energy density of the source. Making the obvious ansatz of spherical symmetry for 
${\bf{\underline h}}$ in the same coordinate system, we find also a Schwarzschild solution with one free constant to be fixed by integrating over the source term, and the pull-over can be set to be constant. Since $g/\underline h =1$ , we see that the integrational constant is just $- M c_W < 0$, where $c_W$ is the determinant of the pull-over. A further symmetry requirement that the asymptotic limit of ${\bf h}$ be just the Minkowski metric ${\bf \eta}$, as is that of ${\bf \underline h}$ and ${\bf g}$, fixes the pull-over to be the identity and $c_W=1$. We then have 
\beqn
g_{tt} &=& - \left( 1-\frac{2M}{r}\right) ~,~ g_{rr} = - 1/g_{tt} ~, \nonumber \\
g_{\theta\theta} &=& r^2 ~,~ g_{\phi\phi}= r^2 \sin^2 \theta \quad,\\
h_{{\underline{tt}}} &=& - \left( 1+\frac{2 M}{r}\right) ~,~ h_{{\underline{rr}}} = - 1/h_{{\underline{tt}}}~,\nonumber \\ h_{{\underline{\theta\theta}}} &=& {r}^2~,~ h_{\underline{\phi\phi}}= {r}^2 \sin^2 \theta 
\quad ,
\eeqn
and $h_{\kappa\nu} = h_{\underline{\kappa\nu}}$. 

One can now compute the connection coefficients according to eq. (\ref{connh}) and obtain the geodesic equations for an $h$-field in this background of a $g$-source. Since ${\bf h}$ is just
a Schwarzschild metric with a negative source one sees e.g. by taking the Newtonian limit that an $h$-particle will
be repelled by the $g$-source. It should be emphasized that in the here presented approach this does not result in
contradictions as those pointed out by Bondi \cite{Bondi} since the $h$-fields do move according to a different
connection and thus the equivalence principle does not apply (see also Appendix B).

In case the radius of the matter source fell below its Schwarzschild radius, and we had
a black hole geometry for ${\bf g}$, the metric ${\bf \underline h}$ would not have a horizon (the $h$-fields are repelled by the source). Note also that this evident symmetry of the above metrics is not as obvious in every coordinate system. For example changing to one of the more well-behaved systems with e.g. in/out-going EF-coordinates will be a nice transformation for the usual metric ${\bf g}$, but 
completely mess up the other metric ${\bf \underline h}$. The reason is just that for the $h$-observer in/out-going means something different. 

As this example shows, the bi-metric model is not causal in the sense that the $h$-field's 
propagation does not need to lie within the lightcone of the standard
fields. The $h$-metric on the manifold thus describes a
global causal structure that in general will be different 
from that of the g-matter. Because of the symmetry between both
however, the $h$-field's propagation is causal as well 
if the properties of a curve's tangent vector are defined
through the respective metric. Physically, there are then two
ways to form closed timelike curves: The one is through an 
interaction between both types of matter which together could 
carry information around a closed timelike curve. Since the interaction 
between both types of matter is mediated only by gravity and very
weak, this would be relevant only for curves going through regions 
where the gravitational interaction is strong. The other possibility 
is through a space-time structure that allows closed timelike curves, 
which is known to be possible in the presence of negative energy 
densities \cite{Morris:1988tu}, even for only small amounts \cite{Visser:2003yf}. 
It remains an open question though whether these  
solutions actually describe natural settings.   

It is further worthwhile to point out that the analysis about stability of the negative mass Schwarzschild black hole worked out in \cite{Gleiser:2006yz} does not apply to the here discussed case, since a perturbation of the metric ${\bf \underline h}$ is related to that of the usual metric ${\bf g}$ via the field equations. Since ${\bf g}$ is stable as usual, there is no reason to expect any instabilities for the metric that the $h$-observer would measure.   

Both types of fields only interact gravitationally, so the $h$-fields constitute 
a kind of very weakly interacting dark matter. Since both kinds of matter repel, one would expect 
the amount of $h$-matter in our vicinity to presently be very small.  
 
\section{Example II: Friedmann-Robertson-Walker}
\label{example2}

We have the usual Friedmann-Robertson-Walker (FRW) metric for ${\bf g}$
\beqn
ds^2 = - dt^2 + \frac{a^2}{1-k r} ( dr^2 + d \Omega^2) \quad,
\eeqn
and make the ansatz for ${\bf \underline h}$
\beqn
ds^2 = - d t^2 + \frac{b^2}{1-k r} ( d r^2 + d \Omega^2) \quad,
\eeqn
where $k = -1,0,+1$. To preserve the symmetry of the FRW metric, we further expect the pull-overs to only 
act on the time coordinate and we have $a^2/(1 - k r) = [P_g({\bf g})]_{\underline{rr}}$ and 
$b^2/(1-k r) = [P_{\underline h}({\bf \underline h})]_{rr}$.  For the sources, we use the notation 
\beqn
T^0_{\;\;0} &=&  \rho ~,~  T^i_{\;\;i} =  p \\
\underline T^{\underline 0}_{\;\;\underline 0} &=& \underline \rho ~,~ \underline T^{\underline i}_{\;\;\underline i} = \underline p \quad.
\eeqn 
(Here, the indices $i, \underline i =1,2,3$, and are not summed over). The first Friedmann equations for both metrics then are
\beqn
\left( \frac{\dot a}{a}\right)^2 &=& \rho - W \bigg( \frac{b}{a} \bigg)^3 \underline \rho - \frac{k}{a^2} \quad,\\
\left( \frac{\dot b}{b}\right)^2 &=& \underline \rho - 
\underline V \bigg(\frac{a}{b} \bigg)^3 \rho - \frac{k}{b^2} \label{friedmannb}\quad,
\eeqn
where a dot indicates a derivative with respect to $t$, and the conservation laws read
\beqn
\partial_t (\rho - W \bigg( \frac{b}{a} \bigg)^3 \underline \rho) + 
3 \frac{\dot a}{a} \left( \rho + p + W \bigg( \frac{b}{a} \bigg)^3 ( \underline \rho + \underline p) \right) = 0 \quad,\\
\partial_{ t} (\underline \rho - \underline V \bigg(\frac{a}{b} \bigg)^3 \rho) + 3 \frac{\dot b}{b} 
\left( \underline \rho + \underline p + \underline V \bigg(\frac{a}{b} \bigg)^3 ( \rho + p) \right) = 0 \quad.
\eeqn
Since the pull-overs here act only on the time-coordinate, the solutions for $a$ and $b$ together with the determinants $\underline V$ and $W$ then give ${\bf h}$ and ${\bf \underline g}$. For example if one wanted to compute the equation of motion for an $h$-photon, one had $h_{tt} = - W^2$ and $h_{rr} = b^2/(1- k r)$ with which one obtains the 
non-metric connection via eqs. (\ref{connh}).  

We will in this present work not attempt to discuss the solutions of these equations in all generality, but instead we consider some
specific cases of interest. In the following, $c_{\underline V}$ and $c_W$ are positive valued constants.
\begin{enumerate}
\item In case $\underline \rho = 0$, $W$ does not appear in the equations. If $\rho$ is a matter field, i.e. $p=0$, then $\underline V = c_{\underline V}$  is a solution; if $\rho$ is a radiation field, i.e. $\rho = 1/3 p$,  then $\underline V = c_{\underline V} a/ b$ is a solution. In this case the $\underline V$ could be absorbed into the metrics (e.g. by changing both into comoving coordinates). One should note that with this choice of sources eq. (\ref{friedmannb}) does not have a solution for $k = 0,1$. The case for $\rho = 0$ is similar.
\item If both $\rho$ and $\underline \rho$ are matter fields then $\underline V = c_{\underline V}$, $W = c_W$ is a solution. \item If both $\rho$ and $\underline \rho$ are radiation fields then $\underline V = c_{\underline V}$, $W = c_{\underline W}$ is also a solution.
\item If $\rho$ is a matter field and $\underline \rho$ is a radiation field then $W = c_W b/a$ and $\underline V= c_{\underline V}$. Similarly, if $\underline \rho$ is a matter field and $\rho$ is a radiation field then $\underline V = c_{\underline V}a/b$ and $W=c_W$. In these cases, it is not possible to set both $W$ and $\underline V$ to be constant.
\item If both sources are cosmological constants we have $W = c_W (b/a)^3$ and $\underline V = c_{\underline V}( a/b)^3$. Note that for certain values of the constants $c_W$ and $c_{\underline V}$ the curvature needs to be negative.
\end{enumerate}
As previously mentioned, the pull-overs are only determined up to constants that have to be specified in the
initial conditions. With a suitable choice of these constants, one can achieve the additional source term
to be negligible. Though this scenario does not seem particularly compelling, we want to point out that for this
reason it is possible to reproduce standard GR up to small corrections.

\section{Consequences and Possible Observables}
\label{disc}

In the previous sections we have studied an extension of GR in whose framework sources with
negative gravitational energy appear in the field equations. These additional $h$-fields interact only gravitationally
with our standard matter, and thus couple only extremely weakly. In this section we want to mention some reasons why
this scenario is interesting and worth further examination.

The model we laid out is purely classical. Nevertheless it is worthwhile to consider the vacuum expectation value
of the stress-energy tensor for quantum fields that are coupled to the classical background. We will assume that 
the field content for both, the $g$-fields and the $h$-fields, is identical such that we have e.g. two copies 
of the Standard Model. The vacuum 
expectation value of these quantum fields is just proportional to the respective metric. Though the constant of proportionality is technically seen divergent, one expects this vacuum energy to be regularized at the Planck scale $m_{\rm p}$. This 
leads one to the well known problem that this vacuum energy density is $\sim m_{\rm p}^4$, and far too large to ever 
allow our universe to form the structures we observe. 

If we consider the vacuum solution in the model with exchange symmetry however, we expect a symmetry between both metrics. In the 
case with the maximal number of space-time symmetries, both would just be the Minkowski metric. We then have $\underline h/g = 1$, and the pull-overs are just the identity.  Since the matter content of both types of fields is identical, this means the source 
terms in eqs. (\ref{fe1b}) and (\ref{fe2b}) cancel identically, no matter how
large their values are. This is a consequence of the additional symmetry. Whether or not this solution is
stable or would run away if the constants did not exactly cancel requires further investigation. Needless to say, the measured value of the cosmological constant is not zero, but at least it is closer to zero than to $m_{\rm p}^4$. 

This bring us to another point to be mentioned, namely the extraction of observables from
the data, e.g. the high-redshift supernovae data or the WMAP results. Underlying the data analysis
to obtain constraints on the parameters in the $\Lambda$CDM-model is the usual GR formalism. 
Unfortunately, some parts of this formalism can not be applied in the model discussed here. For
example, to use cosmological perturbation theory a relevant parameter is the relative size of perturbations 
$\delta \rho/\rho$. Typically, one infers from the CMB measurements the perturbations at 
freeze-out were small $\approx 10^{-5}$. This 
however is the size of perturbations relative to the {\sl observable} (usual) matter density. Since we now 
have an only gravitationally interacting density contribution that is negative, and one further would hope 
for symmetry reasons that both densities are of the same order of magnitude, the total gravitating density 
can be smaller than the observed one. Then, the relative density fluctuations could be larger. Besides 
this, both components of matter repel each other which is an effect usually not present. 

Another feature of the scenario becomes clear from the previously discussed example of the Schwarzschild metric. If there
was a localized source of negative energy, it would act as a gravitational lens - but unlike usual matter 
this would be a diverging lens since it would repel our (usual) photons. Such a lensing event
would typically lower the luminosity of the source, an effect that could potentially add up over distance
if the distribution of such sources is substantial. The detection of a diffractive lensing event could
serve as a smoking gun signal for the here proposed scenario.

\section{Summary}
\label{conc}

We have studied an extension of General Relativity with two metrics, and two sorts of fields. Each field
moves according to the Levi-Cevita connection of one of the metrics. The new sort of fields only interacts
gravitationally with our usual matter. We have coupled these fields to
General Relativity. By requiring the action to be symmetric under exchange of the two metrics, and their fields, 
we obtained a model from which we could derive the equations of motions for the two sorts of fields, as well
as the field equations for both metrics. It turned out that the additional fields can make a contribution to
the gravitational stress-energy tensor with a negative energy density. We argued that this does not imply
a vacuum instability since the kinetic energies are still strictly positive and conserved. 
We further investigated the spherical symmetric example with a source of usual matter, and we found that 
the newly introduced particles would be repelled  by this source. We also derived the Friedmann-equations 
within this scenario and discussed some general properties of possible solutions. Finally, we mentioned some
possible consequences for observables, most importantly a diffractive gravitational lensing effect. 

We hope to have shown that the here proposed bi-metric model with exchange symmetry has interesting
properties, and that it can potentially shed light on some so far unresolved questions in cosmology 
and astrophysics.

\section*{Acknowledgements}

I thank Lee Smolin for helpful discussions, and for reading various drafts of this paper. 
Research at Perimeter Institute for Theoretical Physics is supported in
part by the Government of Canada through NSERC and by the Province of
Ontario through MRI.

\section*{Appendix A}

Contracting Eq(\ref{haga})
\beqn
g_{\mu \nu}  = a_{\mu}^{\;\;\alpha} a_{\nu}^{\;\;\beta} ~ h_{\alpha \beta} \label{hagaA}
\eeqn
with $g^{\mu \kappa}$ yields
\beqn
\delta_{\nu}^{\;\;\kappa}  = a^{\kappa \alpha} a_{\nu}^{\;\;\beta} ~ h_{\alpha \beta} = a^{\kappa \alpha} a^{\lambda \beta} 
~ g_{\lambda \nu} h_{\alpha \beta} \quad.
\eeqn
Taking the variation with use of (\ref{vara}) one obtains
\beqn
a^{\kappa \alpha} a^{\lambda \beta} 
~ g_{\lambda \nu} \delta h_{\alpha \beta} + a^{\kappa \alpha} a^{\lambda \beta} 
~   h_{\alpha \beta}  \delta g_{\lambda \nu} &=& 0 \\
\Leftrightarrow a^{\kappa \alpha} a^{\lambda \beta} 
~ g_{\lambda \nu} \delta h_{\alpha \beta} + g^{\kappa\lambda} \delta g_{\lambda \nu} &=& 0 \quad,
\eeqn
and after contracting with $g_{\kappa \mu}$
\beqn
\delta g_{\mu \nu} = - 
a_{\mu}^{\;\; \alpha} a_{\nu}^{\;\; \beta}  \delta h_{\alpha \beta} \quad. \label{sign1A}
\eeqn
From (\ref{hagaA}) one reads off that the inverse of $a_{\nu}^{\;\;\kappa}$ is
\beqn
[a^{-1}]^{\kappa}_{\;\;\beta} = a^{\kappa \alpha} h_{\alpha \beta} \quad,
\eeqn
where
\beqn
[a^{-1}]^{\kappa}_{\;\; \beta} a_{\nu}^{\;\; \beta} = \delta_\nu^{\;\;\kappa} \quad.
\eeqn
Since $a_{\kappa\nu}$ is symmetric, so is $[a^{-1}]_{\kappa\nu}$ and we also have
\beqn
[a^{-1}]^{\beta}_{\;\; \kappa} a_{\beta}^{\;\; \nu} = \delta^\nu_{\;\;\kappa} \quad. \label{invA}
\eeqn
We then use (\ref{invA}) to bring the $a$'s in (\ref{sign1A}) to the other side 
\beqn
\delta h_{\kappa \lambda} = - [a^{-1}]^\mu_{\;\;\kappa} [a^{-1}]^{\nu}_{\;\;\lambda} \delta g_{\mu \nu} \label{signA} \quad. 
\eeqn
If we consider pulling over one of the indices on $a^{-1}$ with use of $P_{\underline h}$, we obtain
\beqn
[a^{-1}]_{\nu}^{\;\;\kappa} [P_{\underline h}]^{\underline \kappa}_{\;\; \kappa} = a^{\kappa \alpha} h_{\alpha \nu} [P_{\underline h}]^{\underline \kappa}_{\;\;\kappa} \quad.
\eeqn
By putting in a pull-over for the index $\alpha$ and its inverse, and absorbing the pull-overs in the definition 
(\ref{ua}) and (\ref{gauha}) we get
\beqn
[a^{-1}]_{\nu}^{\;\;\kappa} [P_{\underline h}]^{\underline \kappa}_{\;\;\kappa} = a^{\mu \underline \alpha} h_{{\underline{\alpha \kappa}}} = a^{\mu}_{\;\;\underline \kappa} \quad. \label{A2}
\eeqn
The pull-overs are linear, so we have
\beqn
\frac{\delta}{\delta h^{\kappa \lambda}} P_{\underline h}({\underline {\cal L}}) = \left[ P_{\underline h} \left( \frac{\delta {\underline {\cal L}}}{\delta h_{\underline{\kappa \lambda}}}\right) \right]_{\kappa\lambda} = \frac{\delta {\underline {\cal L}}}{\delta h_{\underline{\kappa \lambda}}} [P_{\underline h}]^{\underline \kappa}_{\;\;\kappa} [P_{\underline h}]^{\underline \lambda}_{\;\;\lambda} \quad.
\eeqn
Note that this is not a performance of the variation, but a rewriting of the derivative with the aim to express the
variation of $\underline {\cal L}$ in a way that the symmetry becomes more apparent. We could leave this term in the 
initial form where variation is with respect to $h^{\nu \kappa}$ but we want to make use of the fact that the form of
$\underline {\cal L}$ is symmetric to the usual Lagrangian under exchange of $g_{\kappa \nu}$ with $h_{\underline{\kappa\nu}}$ (not $h_{\kappa\nu}$) and exchange of the respective covariant derivatives.
 
With Eq.(\ref{signA}) and (\ref{A2}) and we then obtain 
\beqn
\frac{\delta}{\delta h^{\kappa \lambda}} P_{\underline h}({\underline {\cal L}}) \delta h^{\kappa \lambda} 
&=& \left[ P_{\underline h} \left( \frac{\delta {\underline {\cal L}}}{\delta h_{\underline{\kappa \lambda}}}\right) \right]_{\kappa\lambda} \delta h^{\kappa \lambda}  \nonumber \\ &=& \frac{\delta {\underline {\cal L}}}{\delta h_{\underline{\kappa \lambda}}} [P_{\underline h}]^{\underline \kappa}_{\;\;\kappa} [P_{\underline h}]^{\underline \lambda}_{\;\;\lambda}\delta h^{\kappa \lambda}  \nonumber \\
&=& - \frac{\delta {\underline {\cal L}}}{\delta h_{\underline{\kappa \lambda}}} [P_{\underline h}]^{\underline \kappa}_{\;\;\kappa} [P_{\underline h}]^{\underline \lambda}_{\;\;\kappa} [a^{-1}]_\mu^{\;\;\kappa} [a^{-1}]_{\nu}^{\;\;\lambda} \delta g^{\mu \nu} \nonumber \\
&=& - \frac{\delta {\underline {\cal L}}}{\delta h_{\underline{\kappa \lambda}}} a_\mu^{\;\;\underline \kappa} a_{\nu}^{\;\;\underline \lambda} \delta g^{\mu \nu} \quad,
\eeqn
with which we return to Eq. (\ref{fe1}).

\section*{Appendix B}

If it wasn't the case that a negative mass particle moved according to a different covariant derivative, one could 
construct the following problem with negative sources in General Relativity: Gravity is a spin-two field. Thus, 
like charges attract and unlike charges repel, and a negative mass particle should be repelled by a positive 
mass source. On the other hand, the negative mass particle moves according to the geodesic equation which does not know anything about the particle's mass - it only knows about the positive source background. Thus, the negative mass
particle should be attracted to the source as all test particles. One would then be lead to conclude a negative mass test particle was attracted {\sl and} repelled likewise which can be used to construct all kinds of nonsense. 

The reason for this confusion is that the use 
of the usual geodesic equation for the negative mass particle is inappropriate which one can understand most easily 
by interpreting the covariant derivative as a coupling to the gravitational field that conserves
the total energy of the particle including the potential energy. For a negative mass test particle that is repelled instead of attracted, the conservation law has to be different since it couples differently to the background. This is similar to the coupling of electrons and positrons to the electric field being mediated by different covariant derivatives.

\end{document}